\documentstyle[preprint,aps,eqsecnum,floats]{revtex}

\input epsf

\newcommand{\beq}{\begin{equation}}
\newcommand{\eeq}{\end{equation}}
\newcommand{\beqa}{\begin{eqnarray}}
\newcommand{\eeqa}{\end{eqnarray}}

\newcommand{\kpnn}{\mbox{$K \to \pi \nu \bar\nu$}}

\newcommand{\sm}{\mbox{Standard Model}}
\newcommand{\apnn}{\mbox{$a_{\pi\nu\bar\nu}$}}
\newcommand{\apks}{\mbox{$a_{\psi K_S}$}}

\newcommand{\susy}{\mbox{supersymmetry}}
\newcommand{\susic}{\mbox{supersymmetric}}
\newcommand{\dmd}{\mbox{$\Delta m_D$}}
\def\npb#1{Nucl.\ Phys.\ {\bf B #1}}
\def\plb#1{Phys.\ Lett.\ {\bf B #1}}
\def\prd#1{Phys.\ Rev.\ {\bf D #1}}
\def\prl#1{Phys.\ Rev.\ Lett. {\bf #1}}

\def\zpc#1{Z.~Phys.\ {\bf C #1}}

\begin{document}

\draft

{\tighten

\preprint{
\vbox{
      \hbox{SLAC-PUB-7690}
      \hbox{WIS-97/32/Nov-PH}
      \hbox{hep-ph/9711215}
      \hbox{November 1997} }}

\renewcommand{\thefootnote}{\fnsymbol{footnote}}

\title{Probing the Flavor and CP Structure of
Supersymmetric Models with $K\to\pi\nu\bar\nu$ Decays
\footnotetext{Research at SLAC supported
by the Department of Energy under contract DE-AC03-76SF00515}}
\author{Yosef Nir${}^{a}$ and Mihir P. Worah${}^{b}$}
\address{ \vbox{\vskip 0.truecm}
${}^a$ Department of Particle Physics, \\
       Weizmann Institute of Science, Rehovot 76100, Israel \\
${}^b$ Stanford Linear Accelerator Center, \\
        Stanford University, Stanford, CA 94309, USA}

\maketitle
\thispagestyle{empty}
\setcounter{page}{0}
\begin{abstract}

We study the implications of various supersymmetric models on the rare
$K^+ \to \pi^+ \nu \bar\nu$ and $K_L \to \pi^0 \nu \bar\nu$ decays.
Although large effects are possible in generic \susic\
models, most of the known \susic\ flavor models lead to negligible
effects. Thus, it is likely that one can get information about CKM
matrix elements from these decays even in the presence of \susy.
Moreover, the possibility of large contributions to \kpnn\ in generic
\susy\ models can be constrained by improved bounds on $D-\bar D$ mixing.
We show that it may be possible to distinguish between different
\susic\ flavor models by combining the information from the \kpnn\ decays
with that from $B-\bar B$ and $D-\bar D$ mixing.

\end{abstract}

} % end tighten

\newpage

%%%%%%%%%%%%%%%%%%%%%%%%%%%%%%%%%%%%%%%%%%%%%%%%%%
%%%%%%%%%%%%%%%%%%%%%%%%%%%%%%%%%%%%%%%%%%%%%%%%
\section{INTRODUCTION}
%%%%%%%%%%%%%%%%%%%%%%%%%%%%%%%%%%%%%%%%%%%%%%%

It has been realized in recent years that the flavor and CP
structure of supersymmetric theories might be very rich.
Measurements of flavor and CP violating processes may become
a sensitive probe of the soft supersymmetry breaking parameters
and, consequently, of the mechanism of dynamical supersymmetry
breaking. For example, various supersymmetric flavor models
give different predictions for the electric dipole moment of
the neutron $d_N$, for CP asymmetries in $B$ decays (e.g.
$a_{\psi K_S}$, the CP asymmetry in $B\to\psi K_S$),
for CP violation in $D-\bar D$ mixing, as well as the $D-\bar D$ mass
difference ($\Delta m_D$) \cite{GNR}.
In this work we study in detail the implications of various
classes of supersymmetric flavor models for the rare \kpnn\ decays.
In particular, BR$(K^+\to\pi^+\nu\bar\nu)$ depends on the flavor
structure of the model, while the ratio
\beq\label{defapnn}
a_{\pi\nu\bar\nu}\equiv{\Gamma(K_L\to\pi^0\nu\bar\nu)\over
\Gamma(K^+\to\pi^+\nu\bar\nu)}
\eeq
depends also on the mechanism of CP violation \cite{Litt}. We show
that combining the predictions for \apnn\ with $a_{\psi K_S}$ and
$\Delta m_D$ allows us to potentially discriminate between the various
\susic\ flavor models.

The \kpnn\ decay is generated by four fermion operators of the form
$\bar s d\bar\nu\nu$. The calculation is essentially clean of hadronic
uncertainties \cite{BuBu,BurasH,MaPa}.
Defining $\theta$ to be the relative
phase between the $s\to d\bar\nu\nu$ decay amplitude and the $K-\bar K$
mixing amplitude, then $a_{\pi\nu\bar\nu}=\sin^2\theta$ \cite{yuv-yos}.
In the \sm, the $K-\bar K$ mixing amplitude is dominated by box diagrams
with intermediate charm and up quarks. The $s \to d\bar\nu\nu$ decay
amplitude
gets significant contributions from both $Z$ penguins and box diagrams.
The dominant contribution in the \sm\ is proportional to $m_t^2$, coming
from diagrams with top quarks in the loops. Had these been the only
important contributions, we would have $a_{\pi\nu\bar\nu}=\sin^2\beta$,
where $\beta$ is the CKM angle,
$\beta\equiv\arg\left[-{V_{cd}V_{cb}^*\over V_{td}V_{tb}^*}\right]$.
However, there is also a smaller but non-negligible charm-loop
contribution proportional to $m_c^2$, with the larger CKM matrix elements
present there compensating for the $m_c^2/m_t^2$ suppression. (Both the
charm and the top quark amplitudes contribute to $K^+\to\pi^+\nu\bar\nu$
rate, whereas the CP violating decay $K_L\to\pi^0\nu\bar\nu$ only gets a
contribution from the dominant top quark amplitude.) The relation between
$a_{\pi\nu\bar\nu}$ and $\sin^2\beta$ is modified, but one still gets
a clean determination of the CKM angle $\beta$ \cite{buras}.

The \kpnn\ decays in \susic\ models have been studied before
\cite{masiero1,giudice1,bigi,konig}.
Our analysis has a different emphasis from the previous analyses:

($i$) Most of these studies have analyzed models of exact
universality \cite{masiero1,giudice1,konig}. (In \cite{bigi}, they
allowed an arbitrary flavor structure; however they did not study the
contribution from chargino penguins, which we find to be the
potentially dominant one.) Recently, various predictive and viable
mechanisms for naturally suppressing the supersymmetric flavor violation
have been suggested which do not assume exact universality. Our main
focus is put on these recent models that have a much richer flavor
structure.

($ii$) All previous studies have only studied the $K^+\to\pi^+\nu\bar\nu$
decay. We study also the $K_L\to\pi^0\nu\bar\nu$ decay. This is
important since the latter is also sensitive to the \susic\ CP violation.
Furthermore, various experimental proposals were recently made to
measure this challenging mode and it is not unlikely that the rates
of both decays will be known in the future.\footnote{Recently, first
evidence for the $K^+\to\pi^+\nu\bar\nu$ decay was presented by the
E787 Collaboration \cite{Adler}.}

($iii$) We emphasize the theoretical strength of combining the
information from \apnn\ and \apks\ \cite{NirLP}. Within the \sm, these
two quantities are strongly correlated because they both depend on
$\beta$ (see Fig. 1).
In most classes of \susic\ flavor models, this relation is
violated. Moreover, looking for deviations in the patterns of flavor
physics like comparing \apnn\ with \apks\ is independent of the hadronic
uncertainties that enter the constraints on $\beta$ and, furthermore, of
the effects that new physics might have on determining the CKM
parameters, unlike the predictions for the rates which sensitively depend
on them.

We find that most of the known \susic\ flavor models have a negligible
effect on \apnn. Therefore, \apnn\ is likely to give us a clean
measurement of $\beta$ even in the presence of supersymmetry. In
contrast, many \susic\ flavor models generally give large contributions
to $B_d-\bar B_d$ mixing, resulting in the fact that we no longer have
$\apks = \sin 2\beta$. While this new contribution may be hard to
detect from just a comparison of \apks\ to the presently allowed range for
$\sin2\beta$ \cite{GNW}, it is likely to be signalled by comparing
to the range that will be allowed by a measurement of \apnn.

We further note that it is possible to generate large corrections to the
\kpnn\ decay rates and to \apnn\ in $R$ parity conserving models
if one allows an arbitrary flavor
structure in the supersymmetric sector. However, in this case, large
contributions to the \kpnn\ decays are often accompanied by large,
detectable contributions to $D-\bar D$ mixing. Thus, improved bounds on
$D-\bar D$ mixing would further constrain the possibility of large
\susic\ contributions to the \kpnn\ decays. (For the future prospects
of searching for $D-\bar D$ mixing, see {\it e.g.} \cite{Liu}.)

%%%%%%%%%%%%%%%%%%%%%%%%%%%%%%%%
%%%%%%%%%%%%%%%%%%%%%%%%%%%%%%%%
\section{Supersymmetric Contributions to $K \to \pi\nu\bar\nu$}

%%%%%%%%%%%%%%%%%%%%%%%%%%%%%%%%
\subsection{General Considerations}

The dominant new contributions to the \kpnn\ decays in supersymmetric
models come from $Z$-mediated penguin diagrams, with supersymmetric
particles inducing the effective $\bar s d Z$ coupling.
Integrating out the $Z$, leads to the
relevant $\bar s d\bar\nu\nu$ four fermion operator.

The analysis is simplified by noting the following points.

1. The effect is always proportional to $SU(2)_L$ breaking. In the
absence of $SU(2)_L$ breaking, the corrections to the
$\overline{s_L} d_L Z$ ($\overline{s_R}d_R Z$) coupling are proportional
to the corrections to the $\overline{s_L}d_L\gamma$
($\overline{s_R}d_R\gamma$) coupling, which vanish at $s=0$, where $s$ is
the four-momentum squared of the intermediate boson. This is the source
of the $~m_t^2$ factor in the dominant \sm\ contribution to \kpnn. In the
supersymmetric framework, wino-higgsino mixing or $\tilde q_L-\tilde q_R$
mixing can provide the necessary $SU(2)_L$ breaking insertion.
Note that the $SU(2)_L$ breaking contributions
generate corrections that do not decouple when the masses of the
fermions in the loop are much larger than $\sqrt{s}$ \cite{ng}.

2. Magnetic moment type couplings of the form $\overline{s_L} d_R Z$
are proportional to the masses of the fermions on the external legs, and
therefore are unimportant.

A closely related calculation is that of the supersymmetric contributions
to the effective $b\bar bZ$ coupling. Theoretically,
both vertices have exactly the same structure (modulo their dependence
on different flavor mixing matrix elements),
one difference being that
the effective $b\bar bZ$ vertex for $R_b$ is evaluated at
$s=M_Z^2$, whereas the $s\bar dZ$ vertex relevant for \kpnn\ is to be
evaluated at $s=0$. Thus, it
may be possible in some models to constrain possible new
contributions to one from the other observable.

These observations lead to the following expectations:

$(i)$ There is a potentially non-negligible contribution from the
charged Higgs -- top loop \cite{buras1}. This contribution is
proportional to the top quark mass squared. It can reach
$\sim 15\%$ of the \sm\ amplitude for $\tan\beta\sim 1$, and a charged
Higgs mass $m_{H_+} \sim 300$ GeV. This contribution can be enhanced
by either smaller values of $\tan\beta$ or a smaller charged Higgs
mass. However, such an enhancement is disallowed by its correlated
contribution to the $\overline{b_L} b_L Z$ vertex resulting in a
large negative correction to $R_b$ \cite{finnell,grant}.
It is also disfavored by
constraints from the $b \to s \gamma$ decay rate \cite{misiak}.
Although both of these constraints can be avoided by invoking
cancellations with chargino-squark diagrams, significant
cancellations are not generic and require fine-tuning.
This contribution scales like $1/\tan^2\beta$ and, consequently,
its significance decreases rapidly for larger values of $\tan\beta$.
Note that the charged Higgs amplitude and the \sm\ top amplitude
are in phase and interfere constructively. Therefore, in the parts
of parameter space where the charged Higgs contribution is
significant and, furthermore, there are no other significant
new contributions, the importance of the charm contribution is
weakened and, consequently, $a_{\pi\nu\bar\nu}$ is predicted to
lie between the \sm\ prediction and $\sin^2\beta$.

$(ii)$ Gluino -- down squark penguins are negligible. To get a
non-decoupling contribution to the $\overline{s_L} d_L Z$ {\em{or}}
$\overline{s_R} d_R Z$ vertex, two $A_d$-insertions are required.%
\footnote{We disagree with the calculation of Ref. \cite{bigi}, where
non-zero effects from gluino penguins with one flavor changing $(RR)$
mass insertion are found.}
Since left-right mixing in the down squark sector is proportional
to down quark masses, the contribution is very small.

$(iii)$ The chargino -- up squark penguins generate only the
$\overline{s_L} d_L Z$ vertex since both the wino coupling and the large
top Yukawa enhanced higgsino coupling (for small $\tan\beta$) are only
to left-handed down-type quarks. The relevant $SU(2)_L$
breaking insertion in the loop involves either $A_t$, namely
$\tilde t_L - \tilde t_R$ mixing, or $\tilde w^+ - \tilde h^+$ mixing.
The size of this contribution depends sensitively on the flavor
structure of the model. In models where the squark mass matrix is
diagonal (up to $\tilde t_L-\tilde t_R$ mixing) in the Super-CKM basis,
the contribution is small, typically $\lesssim 5\%$ of the \sm\
amplitude for super-partner masses $\gtrsim 200$ GeV. Furthermore,
it is in phase with the \sm\ top quark contribution and can partially
cancel the charged Higgs contribution. Allowing super partner masses
to be as small as possible consistent with data could slightly enhance
this contribution.

Introducing flavor violation into the $(RR)$ squark mass-squared
matrix does not enhance the contribution since the higgsino couplings
of the first and second generation squarks are suppressed by factors
of $m_u$ and $m_c$ respectively. However, in supersymmetric models with
misalignment between the $(LL)$ squark and quark mass matrices, large
effects are possible, in principle, with new CP violating phases
(see \cite{konig} for a short discussion of this point).

Thus we learn that:

(a) Without SUSY flavor violation but with light ($\lesssim100\ GeV$)
super partner masses, the SUSY contribution can reach 10\% of
\sm\ and it is always in phase with the \sm\ top penguin.

(b) With SUSY flavor violation in mass-squared matrices for
`left-handed' up-type squarks, the SUSY contribution
could be as large as the \sm\ contribution and with an arbitrary phase.

%%%%%%%%%%%%%%%%%%%%%%%%%%%%%%%%
%%%%%%%%%%%%%%%%%%%%%%%%%%%%%%%%
\subsection{Phenomenological Constraints}

The relevant flavor violating couplings could also affect other FCNC
processes such as neutral meson mixing \cite{BBMR,GGMS,MPR}.
Since, as argued above, a potentially dominant contribution to the
$\kpnn$ decays can come from chargino -- up squark penguins,
we study flavor violation arising from the mass-squared matrix
$m^2_{U_L}$ for the `left-handed' up squarks.
Two important points are: firstly that the off-diagonal entries in
$m^2_{U_L}$ are also constrained by the limits on $m^2_{D_L}$ due to the
relation $m^2_{D_L}= V^{\dagger}m^2_{U_L} V$, where $V$ is the CKM matrix
\cite{MPR} 
\footnote{This bound was overlooked in the original version of this
paper. We thank A. Buras, A. Romanino and L. Silvestrini
for bringing it to our attention.}.
Secondly that the effective second family to first family
($2\to 1$) flavor changing transition in the up-type squark mass matrix
required for the \kpnn\ decay in many cases also leads to large
$D-\bar D$ mixing from box diagrams with gluinos and the up-type squarks.

In our calculations we use the results of \cite{finnell}
(appropriately generalized to allow for arbitrary flavor structure)
to calculate the effective $\bar s d Z$ coupling.
We have checked that our results for the Wilson coefficients of the
$\bar s d\bar\nu\nu$ four fermion operator agree with those presented
in \cite{misiak}.
To get a feel for the size of the flavor violations allowed, we use
the following parameters: $m_0,~M_3,~M_2,~\mu,~A \sim 200-300$ GeV and
$\tan\beta = 2$.%
\footnote{A mass insertion analysis applied
to chargino diagrams requires some care. Besides the obvious
dependence on $\tan\beta$, in special cases, large
off-diagonal mass terms in the squark mass matrix can be cancelled by
factors of CKM matrix elements that are present in the $d_L-\tilde
u_L- \tilde w$ vertex in the Super-KM basis. An example of this
arises in certain U(1) models of alignment \cite{NiSe,LNS}
in which there is a basis where all the mass matrices except the up quark
mass matrix are almost exactly diagonal.}
The interesting diagrams are those where the $SU(2)_L$ breaking is
introduced by higgsino -- wino mixing while the flavor violation is
introduced by one of the following four options:

(i) $(m^2_{U_L})_{12}$ insertion: This gives a large contribution
to $K-\bar K$ mixing by gluino box
diagrams due to its relation to $(m^2_{D_L})_{12}$, 
and one obtains the limit
${|(m^2_{U_L})_{12}|\over\tilde m^2} \lesssim 0.05$. Values close to
this bound lead to large $D-\bar D$ mixing by gluino box diagrams
\cite{NiSe,LNS}. Moreover, one can obtain contributions to the \kpnn\
amplitude of the order of the \sm\ contribution consistent with this
bound. Similarly large contributions to the $K_L \to \mu^+\mu^-$ decay
are obtained. Note that Im$[(m^2_{U_L})_{12}]^2$ is
constrained to be vanishingly small by $\varepsilon_K$.

(ii) $(m^2_{U_L})_{13}$ insertion and $V_{ts}$ factor in the wino
coupling: The strongest constraint is from $B_d-\bar B_d$ mixing,
leading to ${|(m^2_{U_L})_{13}|\over \tilde m^2} \lesssim 0.3$.
This does not contribute to $D-\bar D$ mixing (since
the supersymmetric contribution to the latter involves gluinos).
One obtains supersymmetric contributions to \kpnn\ of order 10\%-20\%
of the \sm\ amplitude.

(iii) $(m^2_{U_L})_{23}$ insertion and $V_{td}$ factor in the wino
coupling: 
This is constrained by its effect on $K-\bar K$ mixing through 
$m^2_{D_L}$, and one obtains 
${|(m^2_{U_L})_{23}|\over \tilde m^2} \lesssim 0.5$.
This again does not
contribute to $D-\bar D$ mixing but can, however, lead to 10\%-20\%
effects in the \kpnn\ decay amplitude.

(iv) $(m^2_{U_L})_{13}(m^2_{U_L})_{23}$ insertion: This gives a large
contribution to $D-\bar D$ mixing. The experimental bound on
$\Delta m_D$ leads to the bound
$\sqrt{{|(m^2_{U_L})_{13}|\over \tilde m^2}
{|(m^2_{U_L})_{23}|\over \tilde m^2}} \lesssim 0.3$.  
We can obtain
contributions to the \kpnn\ amplitude of 30\% the \sm\ contribution
consistent with this bound and those above.
Moreover, $\varepsilon_K$ constrains the relative phases of
$(m^2_{U_L})_{13}$ and  $(m^2_{U_L})_{23}$.

We collect the results above in Table 1. The constraints quoted in this
table are imposed by demanding the supersymmetric contribution to
measured quantities to be less than (a factor of two) the \sm\
contribution and to quantities where only an upper bound exists to be
less than this bound.
Moreover, in a recent publication \cite{BRS}, the effects of large
off-diagonal $(LR)$ entries in the up-type squark mass matrix have
been considered, and shown to result in potentially large
contributions to \kpnn. In the specific models discussed below,
however, these effects are small. 

\begin{table}
\[ \begin{array}{|c|c|c|c|c|}
\hline
{\rm Insertion} & {\rm Constraint} & {\rm Limit}
& {\Delta{m_D}\over2\times10^{-13}\ GeV} &
{A(K^+\to\pi^+\nu\bar\nu)^{\rm SUSY}\over
A(K^+\to\pi^+\nu\bar\nu)^{\rm SM}} \\
\hline
(m^2_{U_L})_{12} & \Delta{m_K} & 0.05 & 0.5 & 0.5 \\
(m^2_{U_L})_{13} & \Delta{m_{B_d}} & 0.3 & 0 & 0.2 \\
(m^2_{U_L})_{23} & \Delta{m_K} & 0.5 & 0 & 0.2 \\
\sqrt{(m^2_{U_L})_{13}(m^2_{U_L})_{23}} & \Delta{m_D} & 0.3 & 1 & 0.3 \\
\hline
\end{array} \]
\vskip 12pt
\caption
{Constraints on the flavor violation in the ``left-handed'' up-type
squark mass-squared matrix, and the largest possible contributions to
$\Delta m_D$ and \kpnn\ consistent with these constraints.
We have used $\tan\beta=2$, a universal scalar mass of 200 GeV,
and chargino and gluino masses in the same range. The quoted limits
are on the ratio between $(m^2_{U_L})_{ij}$ and the typical
supersymmetric mass scale $\tilde m^2$. The bounds from neutral meson
mixing scale like $\times\left({\tilde m\over200\ GeV}\right)$.
${A(K^+\to\pi^+\nu\bar\nu)^{\rm SUSY}}$ scales like
$\times\left({200\ GeV\over\tilde m}\right)^2$.
}
\end{table}

Before concluding this section let us emphasize that the considerations
above generically apply to the coefficient of the
$\overline{s_L}d_L\overline{\nu_L}\nu_L$ four fermion operator. Although
one can trivially obtain from these the corrections to the
$K^+\to\pi^+\nu\bar\nu$ decay rate
({\it i.e.} it is proportional to the absolute value squared),
the rate for the $K_L\to\pi^0\nu\bar\nu$ decay depends also
on the CP structure of the model ({\it i.e.} it is proportional to the
imaginary part squared). In particular, the ratio
$a_{\pi\nu\bar\nu}$, defined in Eq. (\ref{defapnn}),
is sensitive to the supersymmetric CP violation and, in many
cases, only weakly dependent on the flavor violation.

%%%%%%%%%%%%%%%%%%%%%%%%%%%%%%%%
\section{Supersymmetric Flavor Models}

In this section, we apply the above general analysis to specific
classes of models. The classification of models is explained in detail
in ref. \cite{GNR}.

{\it (i) Exact Universality}: At some high energy scale, all squark
masses are universal and the $A$ terms are proportional to the
corresponding Yukawa couplings. Then the Yukawa matrices represent
the only source of flavor (and possibly of CP) violation which
is relevant in low energy physics. There exists the \susic\
analogue of the GIM mechanism which operates in the \sm.
Flavor violations can feed into the soft terms via renormalization
group evolution. The only significant supersymmetric effect is
$\tilde t_L-\tilde t_R$ mixing. This was discussed early in the paper.
The contribution to the \kpnn\ amplitudes could reach $10\%$ for very
light super particle masses. There is no
significant effect on any of the other observables.

{\it (ii) Approximate CP}: The supersymmetric CP problems are solved
if CP is an approximate symmetry broken by a small parameter of
order $10^{-3}$. The flavor structure of this class of models is not
well defined. If we just assume that the flavor violations could
saturate the upper bounds from FCNC processes, then  all of the
observables above could get significant contributions. In particular,
there could be a strong enhancement of BR($K^+\to \pi^+ \nu\bar \nu$).
However, since all CP violation is of order $\varepsilon_K \sim
10^{-3}$, BR($K_L\to \pi^0 \nu\bar \nu$) will not be similarly
enhanced. Instead, we expect $a_{\pi\nu\bar\nu}\sim10^{-3}$.
CP violation in $B$ and $D$ decays is also expected to be of
order $10^{-3}$.

{\it (iii) Abelian H (alignment)}: The squark mass-squared matrices
have a structure, but they have a reason to be approximately
diagonal in the basis set by the quark mass matrix. This is achieved in
models of Abelian horizontal symmetries \cite{NiSe,LNS,NiRa}.
In these models $(m^2_{U_L})_{12}$ is required to be $\sim\theta_C$
(in the super-CKM basis), necessarily leading to large $D-\bar D$ mixing.
However, in processes involving external down-type quarks,
there is a cancellation by
factors of CKM matrix elements that are present in the $d_L-\tilde
u_L- \tilde w$ vertex in the Super-KM basis. If this cancellation
is the weakest consistent with the $\Delta m_K$ constraint,
that is, in the interaction basis
\beq\label{Abe}
{(m^2_{U_L})_{12}\over\tilde m^2} \sim 0.05,
\eeq
then a large contribution to the $K^+\to \pi^+ \nu\bar \nu$ rate
is predicted. This could lead to either enhancement or suppression,
depending on the sign of the supersymmetric amplitude. However, in such a
case, the new contribution is essentially in phase with the $K-\bar K$
mixing amplitude (to satisfy the $\varepsilon_K$ constraint).
Consequently, $K_L\to \pi^0 \nu\bar \nu$ will {\it not} be affected:
it will get contributions from the \sm\ diagrams only.
In all existing models of alignment, the situation is different.
The $\varepsilon_K$ problem is solved by an almost exact cancellation
between the $(m^2_{U_L})_{12}$ insertion and the flavor changing
$d_L-\tilde u_L- \tilde w$ vertex insertion. This requires
${(m^2_{U_L})_{12}\over\tilde m^2} \leq0.004$, but in some explicit
examples the suppression is even stronger ($\sim3\times10^{-5}$).
Under such circumstances, the supersymmetric contributions to the
\kpnn\ rates are negligibly small. Moreover couplings of the third
family via $(13)$ or $(23)$ insertions are too small to have any effect.

{\it (iv) Non-Abelian H (approximate universality)}:
A non-Abelian horizontal symmetry is imposed, where quarks of the
light two families fit into an irreducible doublet. The resulting
splitting among the squarks of these families is very small, leading to
essentially no flavor violation in the $(LL)_{12}$ sector.
The third family supermultiplets are, however, in singlets of the
horizontal symmetry allowing for much larger flavor violating effects.
We have examined several models of this type \cite{BHR,CHM,HaMu}.
We find that, similarly to the models of Abelian horizontal symmetries
discussed above, the mixing between the third generation and the
first two is not of order one but rather of the order of the
corresponding CKM elements:
\beq\label{NonAbe}
{(m^2_{U_L})_{13}\over\tilde m^2} \sim |V_{ub}|,\ \ \
{(m^2_{U_L})_{23}\over\tilde m^2} \sim |V_{cb}|.
\eeq
Such insertions are too small to lead to
significant effects in either \kpnn\ or $D-\bar D$ mixing.

{\it (v) Heavy squarks}: The flavor problems can be solved or,
at least, relaxed if the masses of the first and second generation
squarks $m_i$ are larger than the other soft masses, $m_i^2\sim
100\tilde m^2$ \cite{DKS,DKL,PoTo,CKN}. This does not necessarily lead
to naturalness problems, since these two generations are almost
decoupled from the Higgs sector. A detailed study of the implications
on flavor and CP violation is given in ref. \cite{CKLN}. In the mass
basis, the gluino interaction mixing angles $Z_{ij}^q$ are constrained by
naturalness:
\beq\label{EfSuZ}
|Z_{13}^u| \leq {\rm max}\left({m_{\tilde Q_3}\over\tilde M},|V_{ub}|
\right),\ \ \
|Z_{23}^u| \leq {\rm max}\left({m_{\tilde Q_3}\over\tilde M},|V_{cb}|
\right).
\eeq
The ratio $m_{\tilde Q_3}/\tilde M$ is of order $1/20$ in these models.
The mass of the third generation squarks is of order 1 TeV.
Then diagrams with third generation squarks give only small
contributions. The situation regarding $Z_{12}^u$ is less clear.
Some mechanism to suppress FCNC in the first two generations
(beyond the large squark masses) is necessary in order to
satisfy the $\Delta m_K$ constraint (not to mention the
$\varepsilon_K$ constraint). Even if we assume only mild alignment,
so that $Z_{12}^u\sim\sin\theta_C$ then, for squark mass of order 20 TeV,
the contribution is small. (One can get order 20\% contributions to
\kpnn\ and large $D-\bar D$ mixing if there is not only large mixing
between the first two generation squarks, but also their mass scale is
lower than $\sim 4$ TeV. But then some extra ingredients are required
to explain the smallness of $\varepsilon_K$ in $K-\bar K$ mixing.)

%%%%%%%%%%%%%%%%%%%%%%%%%%%%%%%%
\section{Discussion and Comments}

The situation of \kpnn\ decays in supersymmetric models is
very interesting. It is possible to construct models where there
are significant new contributions to these modes. However, these
models have in general a rather contrived flavor structure
and, moreover, fine-tuned CP violating phases. Consequently,
in most supersymmetric models where the flavor and CP problems
are solved in a {\it natural} way, the \kpnn\ modes get only
small new contributions (of order 5\% or less of the \sm\ amplitude).
In many cases, these new contributions are in phase with the dominant
\sm\ amplitude, the top quark penguin and box
diagrams. Thus, $a_{\pi\nu\bar\nu}$
provides a measurement of the angle $\beta$ of the unitarity triangle to
an accuracy of order 5\% even in the presence of supersymmetry.
The possible exceptions to this statement are models of approximate CP,
where $a_{\pi\nu\bar\nu}\sim10^{-3}$.

The combination of measurements of \kpnn\ decay rates with CP violation
in neutral $B$ decays (and in $D-\bar D$ mixing) may provide a
particularly sensitive probe of supersymmetry. The interesting point is
that, within the \sm, the CP asymmetry in $B\to\psi K_S$, $a_{\psi K_S}$,
and the CP asymmetry in \kpnn, that is $a_{\pi\nu\bar\nu}$ defined in
ref. (\ref{defapnn}), are both devoid of hadronic uncertainties, and
related to the angle $\beta$ of the unitarity triangle. Therefore, the
\sm\ predicts a well-defined relation between the two. Furthermore,
$D-\bar D$ mixing is predicted to be vanishingly small in the \sm. In
some classes of supersymmetric flavor models, these predictions do not
necessarily hold:

(i) Exact universality: The contributions to $B-\bar B$ mixing are
small (up to $O(0.2)$ of the \sm) and with no new phases, so there is
no effect on $a_{\psi K_S}$. Similarly, the contributions to \kpnn\
are small (of $O(0.1)$ of the \sm) and in phase with the top amplitude,
so there is no effect on \apnn. Also
$D-\bar D$ mixing is not affected.

(ii) Approximate CP: CP violating phases are small. Therefore we have
$a_{\psi K_S}\lesssim10^{-3}$, independent of whether SUSY contributions
to the mixing are large or not. Similarly, we have $a_{\pi \nu\bar\nu}
\lesssim10^{-3}$ independent of whether
$K^+\to \pi^+ \nu\bar\nu$ gets a significant contribution or not.

(iii) Alignment: If the relevant squark masses (say the sbottom) are
$\leq O(300\ GeV)$, then the SUSY contribution to $B-\bar B$ mixing can
be $O(1)$. There are arbitrary new CP phases, so $a_{\psi K_S}$
could differ significantly from the standard model.
Moreover, large contributions to $D-\bar D$ mixing are generic in
these models. In existing
models, the contributions to \kpnn\ decays are small and
$a_{\pi\nu\bar\nu}$ is not affected. (However, it may be possible
to construct alignment models with  a larger $(m^2_{U_L})_{12}$, leading
to a large contribution to \kpnn. The $\varepsilon_K$
constraint requires that it is in phase with the \sm\ charm
contribution, thus changing the overall phase of the amplitude, and the
\sm\ expectation for $a_{\pi\nu\bar\nu}$.)

(iv) Approximate universality: The situation with regard to
$a_{\psi K_S}$ is similar to models of alignment. There is
no effect on $a_{\pi\nu\bar\nu}$ or on $D-\bar D$ mixing.

(v) Heavy squarks: Third generation squark masses are $O(1\ TeV)$ and the
first two generations are $O(20\ TeV)$. But the mixing angles in the
gaugino couplings to (s)quarks can be large, so SUSY contributions
to FCNC processes are potentially significant. In particular,
if CP phases are large, then large effects in $a_{\psi K_S}$ are
possible. But if the $\varepsilon_K$ problem is solved
by small phases, then a situation similar to models of approximate
CP might arise. The effect on the \kpnn\ decays is generally small.
%(However, if all the relevant masses are smaller by factor of 5, we could
%get a significant effect on $a_{\pi\nu\bar\nu}$, as well as $D-\bar D$
%mixing.)

Based on this discussion, we list in Table 2 possible flavor physics
signals that could help us to distinguish between the
different SUSY flavor models.
\begin{table}
\[ \begin{array}{|c|c|c|}
\hline
{\rm Model} & \apnn = \apks & \Delta{m_D} > 10^{-14} GeV \\
\hline
{\rm Standard\ Model} & Yes & No \\
{\rm Universality} & Yes & No \\
{\rm Approximate\ CP} & 0 & -- \\
{\rm Alignment} & No & Yes \\
{\rm Approx.\ Universality} & No & No \\
{\rm Heavy\ Squarks} & No & No \\
{\rm Arbitrary} & -- & -- \\
\hline
\end{array} \]
\vskip 12pt
\caption
{Possible outcomes for flavor and CP violation in different models.
By $\apks=\apnn$ we do not mean that they are equal but rather
that they are consistent with the same value of $\beta$, namely
within the allowed range of Fig. 1.
A blank entry implies that there is no specific prediction.}
\end{table}
Thus we see that combining \apnn\ with \apks\ and \dmd, allows us to
partially distinguish between the various \susic\ flavor
models. Moreover, a clean determination of $\beta$ from \apnn\ may
allow us to unambiguously detect new contributions to $B-\bar B$ mixing.

Finally, we stress that in an arbitrary supersymmetric model,
where one just tunes all flavor violating parameters to be consistent
with the phenomenological constraints, it is possible to obtain
deviations from the \sm\ predictions that are different from those
that appear in any of the classes of models discussed above.
In particular, \apnn\ itself can be strongly modified even if
CP is not an approximate symmetry. The absence of large $D-\bar D$
mixing could serve to rule out this possibility.
Also, in this case, the pattern of CP asymmetries
in $B_s$ decays might deviate significantly from both the \sm\ and
the existing \susic\ flavor models. Therefore, measurements of
FCNC and CP violating processes may not only exclude the \sm, but also
require that we examine again the flavor structure of supersymmetry.

\bigskip
{\noindent \bf \it Acknowledgments.}
We thank Yuval Grossman and Jim Wells for useful conversations. Part of
this work was done while MPW was at the Aspen Center for Physics.
YN is supported in part by the United States -- Israel Binational
Science Foundation (BSF), by the Israel Science Foundation,
and by the Minerva Foundation (Munich).

\epsfxsize=356pt
\epsfbox{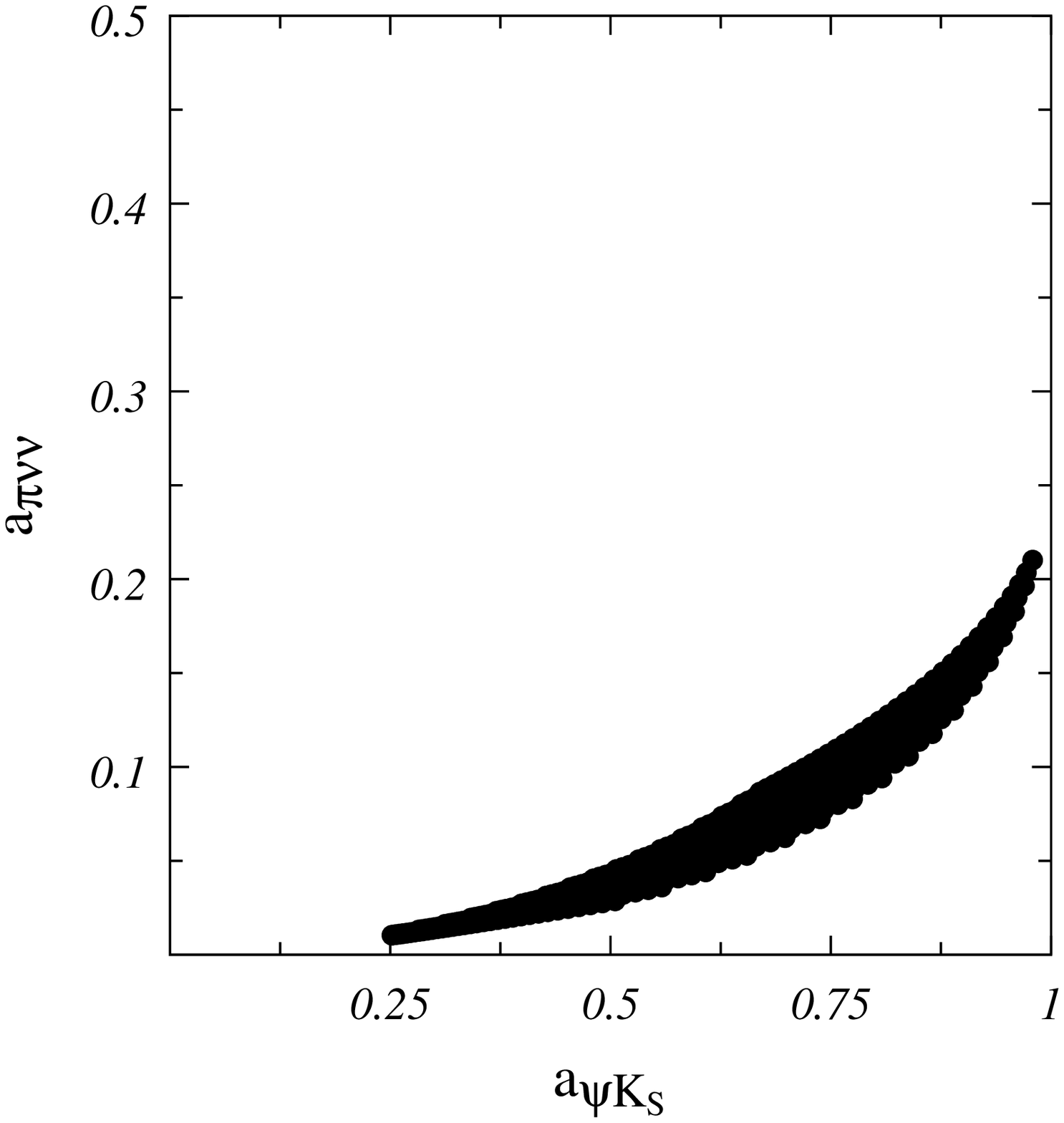}
{\rm Figure 1. }
{The Standard Model allowed region in the \apks - \apnn\ plane.
We have used $-0.25\le\rho\le 0.40$, $0.16\le\eta\le 0.50$ \cite{NirLP}.}


\begin{references}

\bibitem{GNR}
For a recent review, see Y. Grossman, Y. Nir and R. Rattazzi,
 hep-ph/9701231, to appear in the review volume ``Heavy Flavors II",
 eds. A.J. Buras and M. Lindner (World Scientific, Singapore).

\bibitem{Litt}
L.S. Littenberg, \prd{39}, 3322 (1989).

\bibitem{BuBu}
G. Buchalla and A.J. Buras, \plb{333}, 221 (1994), hep-ph/9405259;
\npb{400}, 225 (1993), hep-ph/9707243.

\bibitem{BurasH}
A.J. Buras, \plb{333}, 476 (1994), hep-ph/9405368.

\bibitem{MaPa}
W. Marciano and Z. Parsa, \prd{53}, 1 (1996).

\bibitem{yuv-yos}
Y. Grossman and Y. Nir, \plb{398}, 163 (1997), hep-ph/9701313.

\bibitem{buras}
G. Buchalla and A. Buras, \prd{54}, 6782 (1996), hep-ph/9607447.

\bibitem{masiero1}
S. Bertolini And A. Masiero \plb{174}, 343 (1986).

\bibitem{giudice1}
G. Giudice, \zpc{34}, 57 (1987).

\bibitem{bigi}
I. Bigi and F. Gabbiani, \npb{367}, 3 (1991).

\bibitem{konig}
G. Couture and H. Konig, \zpc{69}, 167 (1995), hep-ph/9503299.

\bibitem{Adler}
S. Adler {\it et al.}, E787 Collaboration, hep-ex/9708031.

\bibitem{NirLP}
Y. Nir, hep-ph/9709301, to appear in the proceedings of the 18th
International Symposium on Lepton Photon Interactions (Hamburg,
July 1997).

\bibitem{GNW}
Y. Grossman, Y. Nir and M. Worah, \plb{407}, 307 (1997), hep-ph/9704287.

\bibitem{Liu}
T. Liu, hep-ph/9508415, a talk presented at the $\tau$-charm Factory
Workshop (Argonne, June 1995).

\bibitem{ng}
J. Liu and D. Ng, \plb{342}, 262 (1995), hep-ph/9408375;
D. Comelli and J. Silva, \prd{54}, 1176 (1996), hep-ph/9603221.

\bibitem{buras1}
A. Buras {\it et al.}, \npb{337}, 284 (1990).

\bibitem{finnell}
M. Boulware and D. Finnell, \prd{44}, 2054 (1991).

\bibitem{grant}
A. Grant, \prd{51}, 207 (1995), hep-ph/9410267.

\bibitem{misiak}
P. Cho, M. Misiak and D. Wyler, \prd{54}, 3329 (1996), hep-ph/9601360.

\bibitem{BBMR}
S. Bertolini, F. Borzumati, A. Masiero and G. Ridolfi,
\npb{353}, 591 (1991).

\bibitem{GGMS}
For a recent study, see F. Gabbiani, E. Gabrielli, A. Masiero
and L. Silvestrini, \npb{477}, 321 (1996), hep-ph/9604387.

\bibitem{MPR}
M. Misiak, S. Pokorski and J. Rosiek,
hep-ph/9703442, to appear in the review volume ``Heavy Flavors II",
 eds. A.J. Buras and M. Lindner (World Scientific, Singapore).

\bibitem{BRS}
A. Buras, A. Romanino and L. Silvestrini, hep-ph/9712398.

\bibitem{NiSe}Y. Nir and N. Seiberg, \plb{309}, 337 (1993),
hep-ph/9304307.

\bibitem{LNS}M. Leurer, Y. Nir and N. Seiberg, \npb{420}, 468 (1994),
hep-ph/9310320.

\bibitem{NiRa}Y. Nir and R. Rattazzi, \plb{382}, 363 (1996),
hep-ph/9603233.

\bibitem{BHR}R. Barbieri, L.J. Hall and A. Romanino,
 \plb{401}, 47 (1997), hep-ph/9702315.

\bibitem{CHM}C.D. Carone, L.J. Hall and T. Moroi, hep-ph/9705383.

\bibitem{HaMu}L.J. Hall and H. Murayama, \prl{75}, 3985 (1995),
 hep-ph/9508296.

\bibitem{DKS}M. Dine, A. Kagan and S. Samuel, \plb{243}, 250 (1990).

\bibitem{DKL}M. Dine, A. Kagan and R.G. Leigh, \prd{48}, 4269 (1993),
hep-ph/9304299.

\bibitem{PoTo}A. Pomarol and D. Tommasini, \npb{466}, 3 (1996),
hep-ph/9507462.

\bibitem{CKN}A.G. Cohen, D.B. Kaplan and A.E. Nelson,
 \plb{388}, 588 (1996), hep-ph/9607394.

\bibitem{CKLN}A.G. Cohen, D.B. Kaplan, F. Lepeintre and A.E. Nelson,
 \prl{78}, 2300 (1997), hep-ph/9610252.

%\bibitem{NeWr}A.E. Nelson and D. Wright, hep-ph/9702359.


\end{references}
\end{document}